# Coupled Magnetic Resonator Optical Waveguides

## Hui Liu, and Shi-ning Zhu

*National Laboratory of Solid State Microstructures, Department of Physics, Nanjing University, People's Republic of China*

Email: liuhui@nju.edu.cn, zhusn@nju.edu.cn; URL: http://dsl.nju.edu.cn/mpp

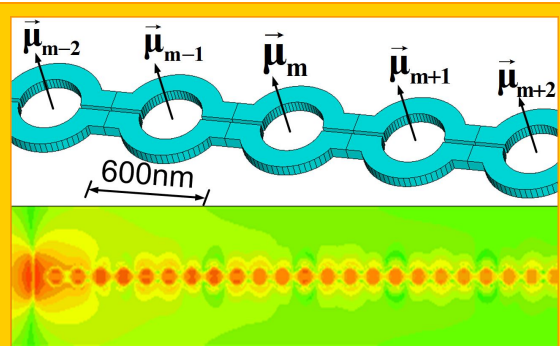

**Abstract** Optical resonators are important devices that control the properties of light and manipulate light-matter interaction. Various optical resonators are designed and fabricated using different techniques. For example, in coupled resonator optical waveguides, light energy is transported to other resonators through near-field coupling. In recent years, magnetic optical resonators based on LC resonance have been realized in several metallic microstructures. Such devices possess stronger local resonance and lower radiation loss compared with electric optical resonators. This study provides an overall introduction on the latest progress in coupled magnetic resonator optical waveguide (CMROW). Various waveguides composed of different magnetic resonators are presented and Lagrangian formalism is used to describe the CMROW. Moreover, several interesting properties of CMROW, such as abnormal dispersions and slow light effects, are discussed and CMROW applications in nonlinear and quantum optics are shown. Future novel nanophotonic devices can be developed using CMROW.

1. Introduction
2. Various periodic CMROW
   2.1 Magneto-inductive waveguide
   2.2 Periodic split-ring resonator chain
   2.3 Periodic split-hole resonator chain
   2.4 Periodic nanosphere chain on slab
   2.5 Periodic nanosandwich chain

**3. Aperiodic CMROW**

**4. Nonlinear CMROW**

**5. Quantum CMROW**

**6. Outlook**

**1. Introduction**

Coupled resonator optical waveguide (CROW) is used to accommodate light propagation in a preferred manner because of the coupling between adjacent resonators [1]. Various dielectric microresonators that constitute CROW, such as microspheres, microdiscs [2], and photonic crystal microcavities [3-5], have been reported. Light propagates in CROW through the near-field coupling between resonators, and the dispersion of wave vectors as well as group velocities can be tuned by changing the coupling process. Therefore, CROW can be used to obtain slow light effects and optical buffers [3, 6, 7] and enhance light-matter interaction, making it suitable for nonlinear and quantum optical processes [8-10].

Researchers have used various techniques to shrink the size of CROW and produce an integrated photonic chip. The size of a dielectric resonator cannot be smaller than half a wavelength because of diffraction limit. On the other hand, the optical properties of plasmonic structures have been widely investigated in the past two decades, concomitant with the remarkable progress in various techniques for nanomanufacturing and chemical fabrication. Plasmon materials have the ability to manipulate photons in the subwavelength scale, making them applicable in many important applications, such as optical information, nonlinear optics, and biosensors, among others. In recent years, the coupling effects among plasmonic nanostructures have increasingly attracted the interests of researchers [11]. Various coupling processes converge into plasmon systems, which behave like chemical molecules or condensed matters and have various complex optical properties. A CROW created from plasmonic resonators, such as metallic nanoparticles, have been proposed to reduce the sizes of optical devices below the diffraction limit [12-15]. An array of closely spaced metal nanoparticles coherently guides the electromagnetic (EM) energy via near-field coupling. Metal particles are known to support the collective electronic excitation of surface plasmon (SP) with resonance frequencies depending on the particle size and shape. Metal nanoparticles with absorption cross-section far exceeding their geometrical sizes exhibit strong light absorption because of SP resonance. Thus, metal nanostructures efficiently convert EM energy into oscillatory electron motion, which is a necessary condition for the strong coupling of light into waveguiding structures.

The magnetic plasmon (MP) resonator is another novel design that widely aroused research interests. In 1999 [16], Pendry reported that nonmagnetic metallic element double split ring resonators (DSRR), with a size below the diffraction limit, exhibits a strong magnetic response and behaves like an effective negative permeability material. Although DSRR systems do not contain free magnetic poles, the excitation of displacement currents in the DSRR results in the induction of a magnetic dipole moment that is somehow similar to a bar magnet. Analogous to the SP resonance in metal nanoparticles, an effective media made of DSRRs can support resonant MP oscillations at GHz [16-18] and THz frequencies [19-21]. Such systems can be combined with an electric response and characterized by negative permittivity to develop metamaterials with negative indices of refraction [17, 18].

According to the classical electrodynamics theory [22], the radiation loss of a magnetic dipole is substantially lower than that of an electric dipole of a similar size. Thus, the use of a coupling magnetic resonators optical waveguide (CMROW) to guide EM energy over long distances has great potential for direct applications in novel sub-diffraction-limited transmission lines without significant radiation loss. Furthermore, near-field coupling interactions between magnetic resonators, such as electric field coupling, magnetic field coupling, and exchange current coupling, are quite complicated. Exchange current coupling, which is stronger than the other two coupling interactions, can introduce broader dispersion band and more efficient energy transfer.

This study provides an overall introduction on the recent developments in CMROW. In figure 1, we show different kinds of CMROW which will be introduced in this paper. Section 2 introduces periodic CMROW structures that are composed of various magnetic resonators, such as split-ring resonator (SRR) chains, slit-hole resonator (SHR) chains, nanosphere chains on slab, and nanosandwich chains. Aperiodic CMROW is then described in section 3, followed by a presentation of nonlinear CMROW in section 4. Afterwards, recent progresses in quantum CMROW are introduced in section 5. Finally, an outlook that predicts possible future developments in CMROW is presented in section 6.

**2. Various periodic CMROW**

2.1 Magneto-inductive waveguide

MP resonance is applied to a 1D sub-wavelength waveguide in the microwave range [23-25]. Shamonina et al. proposed a propagation of waves supported by capacitively loaded loops by using a circuit model in which each loop is coupled magnetically to a number of other loops [23]. The waves are referred to as magneto-inductive (MI) waves because the coupling is caused by induced voltages. MI waves that propagate on 1D lines may exhibit both forward and backward waves depending on whether the loops are arranged in an axial or planar configuration, which are shown in figure 2 (a-b). Moreover, band broadening can be obtained because of the excitation of MI waves, and the bandwidth changes dramatically as the coupling coefficient between the resonators is varied [26]. A kind of polariton mode can be formed through the interaction of electromagnetic and MI waves, resulting in a tenability of the range where the magnetic permeability μ becomes negative [25]. In a biperiodic chain of magnetic resonators, the dispersion of the MI wave is split into two branches that are analogous to acoustic waves in solids and it can be used to obtain specified dispersion properties [27, 28]. In addition, electro-inductive (EI) waves were also reported to be in the microwave range [29]. Furthermore, the coupling may either be magnetic or electric depending on the relative orientation of the resonators, causing the coupling constant between resonators to become complex and consequently leading to even more complicated dispersion [30]. Many microwave devices based on MI waves, such as MI waveguides [31], broadband phase shifters [32], parametric amplifiers [33], and pixel-to-pixel sub-wavelength imagers [34, 35], have been proposed.

2.2 Periodic split-ring resonator chain

The ohmic loss inside metallic structures is much higher in the optical range than in the microwave range. The MI coupling between the elements is insufficiently strong to transfer energy efficiently. The exchange current interaction between two connected SRRs [36], which is much stronger than the corresponding MI coupling, has been proposed to improve the properties of the guided MP wave.

Figure 3(a) shows a design of a single split-ring resonator (SSRR) characterized by two half-space metal loops with their tails adjacent to their ends; the gap between the tails plays the role of a capacitor. For simplicity, the SSRR in the analysis was viewed as an ideal LC circuit composed of a magnetic loop (corresponding to the metal ring) with inductance L and a capacitor with capacitance C (corresponding to the gap). In general, an LC resonator is mathematically equivalent to a classic mechanical resonator and can be described by the Lagrangian formalism of an oscillating resonator [37]. If the total charge Q accumulated in the slit is defined as a generalized coordinate, the Lagrangian formalism corresponding to a SSRR can be written as follows:

$$\text{Lag} = \frac{L}{2}\dot{Q}^2 - \frac{1}{C}Q^2 \qquad (1)$$

where $\dot{Q}$ is the induced current, $L\dot{Q}^2/2$ relates to the kinetic energy of the oscillations, and $Q^2/C$ is the electrostatic energy stored in the SSRR's gaps (in figure 4 (b), the total capacitor of two cascaded gaps is C/2). Solving the Euler-Lagrange equation, $\frac{d}{dt}\left(\frac{\partial \text{Lag}}{\partial \dot{Q}}\right) - \frac{\partial \text{Lag}}{\partial Q} = 0$, the resonance frequency of the structure is known to be $\omega_0 = \sqrt{2}/\sqrt{LC}$.

The magnetic moment of the SSRR originates from the oscillatory behavior of the currents induced in the resonator. Magnetic response excitation in a system of SSRRs fabricated on a planar substrate results in the induction of magnetic dipole moments that are perpendicular to the substrate plane, as shown in figure 3(b). Parallel dipoles are characterized by small spatial field overlaps, and thus, the MI interactions between them are expected to be weak. Thus, the SSRRs were physically connected with one another to substantially increase the coupling between the dipoles, as shown in figure 3(b). The contact between the rings serves as the "bond" for conduction current to flow from one SSRR to another. Thus, the proposed system interacts directly through the exchange of conduction current in addition to MI coupling. Such type of coupling is somewhat similar to the electron exchange interaction between two magnetic atoms in a ferromagnetic material [38]. The introduction of a second SSRR, as shown in figure 3(b), results in the splitting of the MP resonance because of the interaction. The splitting of the MP resonance can also be described by the Lagrangian formalism above. If

$Q_m$ is the total oscillation charge in the *m*-th SSRR (m = 1, 2), L is the induction of the ring, and C is the capacitance of the gap, then the Lagrangian formalism of the coupled system can be written as follows:

$$\text{Lag} = \frac{1}{2}L\left(\dot{Q}_1^2 + \dot{Q}_2^2\right) - \frac{1}{2C}\left(Q_1^2 + Q_2^2\right) + M\dot{Q}_1\dot{Q}_2 - \frac{1}{4C}(Q_1 - Q_2)^2, \qquad (2)$$

where the first two terms correspond to the energy stored in the inductors and the end capacitors, respectively. The interaction term $M\dot{Q}_1\dot{Q}_2$ is caused by the magneto-inductive coupling, while the interaction term $\frac{1}{4C}(Q_1 - Q_2)^2$ comes from the exchange current interaction through the connected gaps between two SSRRs. In our other work of two coupled SSRRs [39], there is no such interaction term as the two SSRRs are separated without exchange current between them. Introducing the ohmic dissipation, $R = \frac{1}{2}\gamma\left(\dot{Q}_1^2 + \dot{Q}_2^2\right)$, and substituting Equation (2) in the Euler-Lagrange equations yield the following:

$$\frac{d}{dt}\left(\frac{\partial \text{Lag}}{\partial \dot{Q}_m}\right) - \frac{\partial \text{Lag}}{\partial Q_m} = -\frac{\partial R}{\partial \dot{Q}_m}, \quad m = 1, 2 \qquad (3)$$

Then, coupled equations for the magnetic moments $\mu_m = A\dot{Q}_m$ (where A is a constant related to the area of SSRR and its geometry) can be obtained as follows:

$$\ddot{\mu}_1 + \omega_0^2\mu_1 + \Gamma\dot{\mu}_1 = \frac{1}{2}\kappa_1\omega_0^2(\mu_1 + \mu_2) - \kappa_2\ddot{\mu}_2$$
$$\ddot{\mu}_2 + \omega_0^2\mu_2 + \Gamma\dot{\mu}_2 = \frac{1}{2}\kappa_1\omega_0^2(\mu_1 + \mu_2) - \kappa_2\ddot{\mu}_1 \qquad (4)$$

where $\omega_0^2 = 2/(LC)$ and $\Gamma = \gamma/L$ are the degenerated MP mode eigenfrequency and bandwidth, respectively. The electromagnetic coupling between the resonators is governed by two separate mechanisms. The first term at the right side of Equation (4) corresponds to the interaction caused by the exchange of conduction current, whereas the second term represents the MI contribution. The coupling coefficients are related to the equivalent circuit characteristics of the SSRR. For instance, $\kappa_2 = M/L$ depends on the SSRR's mutual and self-inductance and for an ideal circuit $\kappa_1 = 1/2$. Equation (4) yields solutions in the form of damped harmonic oscillation $\mu = \mu_{i0}\exp(-\frac{1}{2}\Gamma_i t + i\omega_i t)$, where the index $i = 1, 2$ specifies the MP mode. Using $\Gamma/2\omega_0 \ll 1$, the system eigenfrequencies of $\omega_1 = \omega_0\sqrt{(1-\kappa_1)/(1+\kappa_2)}$ and $\omega_2 = \omega_0/\sqrt{1-\kappa_2}$ can be estimated from Equation (4). The high-frequency (anti-symmetric) mode $\omega_2$ yields $\mu_1 = -\mu_2$ and makes the exchange current interaction term in equation (4) negligible. Consequently, the observed frequency shift $\Delta\omega_2 = |\omega_2 - \omega_0|$ is predominantly caused by the magneto-inductive coupling between the SSRRs. This phenomenon is depicted in figure 3(c), where the local current density inside the resonators is plotted. Two distinctive current loops that are closed through a displacement current at the resonator tails are formed, and no conduction current is shared between the SSRRs. On the other hand, the low-frequency (symmetric) MP mode $\omega_1$ yields $\mu_1 = \mu_2$ and both exchanges of conduction current and magneto-inductive interactions contribute to the frequency shift $\Delta\omega_1 = |\omega_1 - \omega_0|$. Figure 3(d) shows the unimpeded flow of current between the SSRRs. Comparisons between the frequency shifts

$\Delta\omega_1 \gg \Delta\omega_2$, and the absolute values of the coupling constants $\kappa_1 \gg \kappa_2$ show that the exchange of conduction current is the dominant coupling mechanism for the proposed SSRRs system.

The magnetic dipole model described above can also be applied to investigate a finite or infinite chain of connected SSRRs (see figure 4). Thus, if a magnetic dipole $\mu_m$ is assigned to each resonator and only the nearest neighbor interactions are considered, then the Lagrangian and the dissipation function of the system can be written as follows:

$$\text{Lag} = \sum_m \left( \frac{1}{2} L \dot{Q}_m^2 - \frac{1}{4C}(Q_m - Q_{m+1})^2 + M \dot{Q}_m \dot{Q}_{m+1} \right)$$
$$R = \sum_m \frac{1}{2} \gamma \dot{Q}_m^2 \quad (5)$$

Substituting Equation (5) into the Euler-Lagrange equations yields the following equations of motion for the magnetic dipoles:

$$\ddot{\mu}_m + \omega_0^2 \mu_m + \Gamma \dot{\mu}_m = \frac{1}{2} \kappa_1 \omega_0^2 (\mu_{m-1} + 2\mu_m + \mu_{m+1}) - \kappa_2 (\ddot{\mu}_{m-1} + \ddot{\mu}_{m+1}) \quad (6)$$

The general solution of Equation (6) corresponds to an attenuated MP wave $\mu_m = \mu_0 \exp(-m\alpha d)\exp(i\omega t - imkd)$, where $\omega$ and $k$ are the angular frequency and wave vector, respectively, $\alpha$ is the attenuation per unit length, and d is the SSRR's size. By substituting $\mu_m(t)$ into Equation (6) and working in a small damping approximation ($\alpha d \ll 1$), the simplified MP relationships for dispersion can be obtained as follows:

$$\omega^2(k) = \omega_0^2 \frac{1 - \kappa_1[1 + \cos(kd)]}{1 + 2\kappa_2 \cos(kd)} \quad (7)$$

The range of applicability and the overall accuracy of the predicted relationships in figure 4 were compared to the finite-difference time-domain (FDTD) results for finite chain SSRRs. In contrast with the electric plasmon (EP) polariton in linear chain nano-sized metal particles [12-14], where both transverse and longitudinal modes exist, the MP is exclusively a transversal wave that is manifested by a single dispersion curve (represented by a black solid line in figure 4(c)), which covers a broad frequency range $\omega \in (0, \omega_c)$ with a cutoff frequency $\omega_c$. Here, the cutoff frequency $\omega_c$ is the maximum value of the excitation frequency for magnetic plasmon modes in CMROW. The precise contribution of each coupling mechanism in the MP dispersion can be investigated using Equation (7). Exclusion of the magneto-inductive term results in a slight decrease in cutoff frequency $\omega_c \to \omega_0$ (represented by a blue dashed curve in figure 4(c)). On the other hand, if the SSRRs interact only through the MI force, the propagating band shrinks to a very narrow range of frequencies $\Delta\omega \cong 2\omega_0 \kappa_2$ centered around $\omega_0$ (red dotted curve in figure 4(c)). Relatively short bandwidths are characteristics of EP [14] and follow the rapid fall of the MI force with distance. Strong wave dissipation is one of the major obstacles for the utilization of surface plasmons in optical devices. The sub-diffraction-sized MP transmission line promises a considerable improvement in wave transmission. The attenuation of most propagation bands remains constant at relatively low value. The propagation length of MP wave is about 15.4 μm at an incident frequency $\hbar\omega = 0.3\text{eV}$. Here, the loss mainly comes from the internal ohmic loss of metal material.

Magnetic resonance coupling between connected SRRs and MP excitations in other types of connected SRR chains have also been investigated [40]. By changing the connection configuration, the chain provides two kinds of MP bands formed by the collective magnetic resonance in SRRs. Two kinds of configurations of SRRs are proposed called homo-connection (slits at same side) and hetero-connection(slits at opposite sides), as schematically shown in figures 5(a) and 4(b) respectively. Based on the extracted dispersion properties of MPs, the forward and backward characteristics of the guided waves are well exhibited and corresponds to the homo and hetero-connected chains, as shown in figures 5(c) and 5(d), respectively. The revealed MP waves both had wide bandwidths starting from the zero frequency because of conductive coupling. These results are suggested to provide instructions for creating new kinds of subwavelength waveguides. The reversed dispersion properties also can be explained by extending the coupled LC-circuit theory. The reversal of the dispersion is mainly come from the alternation of the electroinductive coupling due the change of the slits configuration. The conductive item attributing from the current exchanges is an important factor to build such a wide MP band, which does not exist in the coupling between the nanoparticles, nano-sandwiches, or some other discrete resonators. The retrieved dispersion maps (not shown here) show they are almost the same within the same frequency range as we concerned here and exhibit an SP wave characteristic that rather different from results of these CMROW formed by SRR chains. At this point, our study provides another method to construct subwavelength CMROWs with wide band that accommodating the MP wave propagation with in a preferred characteristics.

2.3 Periodic split-hole resonator chain

In general, MP resonance frequency increases linearly with decreasing overall SRR size. However, the saturation of the magnetic response of SRR at high frequencies prevents it from achieving high-frequency operations. In addition, the complicated shape and narrow gap of SRRs make experiments very challenging. The SHR [41] is considered as a good alternative for making sub-wavelength waveguides because of its simple structure and high working-frequency regime.

Figure 6(a) shows the designed SHR structure based on the design idea proposed by reference [29]. The designed SHR structure comprises two parts, namely, a nano-hole near the edge of a semi-infinite golden film and a slit that links the hole with the edge. The geometric parameters of the designed SHR structure are also provided in figure 6(a). Compared with SRR, SHR is easier to fabricate and contains a resonance frequency that can reach the infrared range. In the simulations, a well-pronounced resonance mode wherein the electric field is confined within the slit was observed, and the magnetic field was concentrated inside the nano-hole. The SHR can be seen as an equivalent LC circuit with the nano-hole as a conductor and the slit as a capacitor. The induced resonance current in the LC circuit was also obtained in the simulations. The current was only observed at a thin layer (thickness of approximately 30 nm) around the nano-hole because of the skin effect in the metal material. The whole SHR structure is seen as a magnetic dipole when the oscillation current is induced by an external wave at resonance frequency. A semi-analytic theory based on Lagrangian formalism was used to describe the oscillation of the magnetic dipole. If $Q$ is the total oscillation charge in the SHR, L is the inductor of the nano-hole, and C is the capacitance of the slit, then the Largangian equation of the system can be written as follows: $\text{Lag} = \frac{L\dot{Q}^2}{2} - \frac{Q^2}{2C}$. Based on the Euler-Lagrange equation $\frac{d}{dt}\left(\frac{\partial \text{Lag}}{\partial \dot{Q}}\right) - \frac{\partial \text{Lag}}{\partial Q} = 0$, the SHR oscillation equation can be obtained as follows: $\ddot{Q} + \frac{1}{LC}Q = 0$. If the SHR is defined as a single magnetic dipole given by $\mu = \dot{Q} \cdot S$, where S is the circular area of the SHR, then $\ddot{\mu} + \omega_0^2 \mu = 0$, where $\omega_0^2 = 1/(LC)$ is the resonance frequency of the SHR.

Based on the SHR described above, a 1D chain of magnetic resonators can be formed by connecting such a structure one by one. In our previous work, a monatomic chain of SRRs was proposed and the MP mode was found in such system [36]. However, the dispersion relation curve of the monatomic chain of SRRs lied below the light line. Moreover, at a given photon energy, the wave vector was not conserved when the photon was transformed into the MP mode. The MP mode was not excited using a far-field incident wave, and the EM energy was not radiated out from the chain. Therefore, it can be concluded that the MP mode in a monatomic chain cannot lead to extraordinary optical transmission (EOT), contrary to what was expected. A diatomic SHR chain was designed and presented to satisfy the wave vector matching conditions, as shown in figure 6(b). As can be seen in figure 6(b), the unit cell of the proposed chain was composed of two SHRs with different geometric sizes. The Lagrangian equation for the infinite diatomic SHR chains can be expressed as follows:

$$\text{Lag} = \sum_m \left( \frac{L_1 \dot{Q}_m^2}{2} + \frac{L_2 \dot{q}_m^2}{2} - \frac{(Q_m - q_{m-1})^2}{2C} - \frac{(Q_m - q_m)^2}{2C} \right) \qquad (8)$$

where the oscillating charges in the m-th unit cell are defined as $Q_m$ for the bigger SHR with an inductor $L_1$ and as $q_m$ for the smaller SHR with an inductor $L_2$ (m = 0, ±1, ±2, ±3, …). The two corresponding magnetic dipoles, $U_m$ and $\mu_m$, are defined as $U_m = \dot{Q}_m \cdot S$ and $\mu_m = \dot{q}_m \cdot s$, where S and s are the areas of the bigger and smaller SHRs, respectively. Based on the Euler-Lagrange equations $\frac{d}{dt}\left(\frac{\partial \text{Lag}}{\partial \dot{U}_m}\right) - \frac{\partial \text{Lag}}{\partial U_m} = 0$ and $\frac{d}{dt}\left(\frac{\partial \text{Lag}}{\partial \dot{\mu}_m}\right) - \frac{\partial \text{Lag}}{\partial \mu_m} = 0$ (m=0, ±1, ±2, ±3, …), the oscillation equations of the m-th bigger and smaller SHRs can be obtained as follows:

$$\begin{cases} \ddot{U}_m + \omega_1^2 \cdot (2U_m - \mu_m - \mu_{m-1}) = 0 \\ \ddot{\mu}_m + \omega_2^2 \cdot (2\mu_m - U_m - U_{m+1}) = 0 \end{cases} \qquad (9)$$

where $\omega_1 = 1/\sqrt{L_1 C}$ and $\omega_2 = 1/\sqrt{L_2 C}$. A general solution to Equation (9) in the form of the MP wave can be obtained as follows:

$$\begin{cases} U_m = U_0 \cdot \exp(i(\omega t - k \cdot md)) \\ \mu_m = \mu_0 \cdot \exp(i(\omega t - k \cdot (md + d/2))) \end{cases} \qquad (10)$$

where $\omega$ is the angular frequency, $k$ is the wave vector, $U_0$ and $\mu_0$ are the initial values of the magnetic dipole moment at $m = 0$, and $d = 650\,nm$ is the period of the chain. By substituting Equation (10) into Equation (9) and then solving the eigenequations for $U_0$ and $\mu_0$, the MP dispersions can be obtained as follows:

$$\omega_\pm^2 = \left(\omega_1^2 + \omega_2^2\right) \pm \sqrt{\left(\omega_1^4 + \omega_2^4\right) + 2\omega_1^2\omega_2^2 \cos(kd)} \qquad (11)$$

The dispersion relations are numerically depicted as two solid black curves in figure 7(a). The diatomic chain contains two separate dispersion branches, namely, the upper branch $\omega_+(k)$ and lower branch $\omega_-(k)$, the m-th unit cells of which have different resonant manners. The simulated results show that $U_m$ and $\mu_m$ oscillate in the same phase in the lower branch $\omega_-(k)$ and oscillate in the anti-phase in the upper branch $\omega_+(k)$. Using the analogy of the diatomic model of crystal lattice wave [42], the upper curve $\omega_+(k)$ can be referred to as the optical branch and the lower curve $\omega_-(k)$ as the acoustic branch. Compared with the monatomic chain [36], which only possesses the acoustic dispersion branch, the diatomic chain possesses the optical branch as well. The light line in free space is represented by a blue dotted straight line in figure 7(b) ($\omega = ck_0$). The intersection of the upper optical branch with the light line was exciting to observe, and the major part of the curve lied on the left side of the intersection point. For an oblique incident plane wave, the resonant excitation of the MP modes can be achieved under the wave vector matching condition as follows:

$$k = k_0 \sin\theta \qquad (12)$$

where $\theta$ is the incident angle, as denoted in figure 7(a). The dependence of resonance excitation frequency on the incident angle can be solved numerically by combining Equations (11) and (12), as shown by the white line in figure 7(b). The MP mode for a perpendicular incident wave ($\theta = 0^0$) is excited at the frequency of 1.11 eV. At the crossing point of the optical branch curve and the blue line in figure 7(a), the MP mode was excited by a plane wave propagating along the metal surface ($\theta = 90^0$), with the corresponding frequency of 0.924 eV. Thus, the MP mode had an excitation frequency range of 0.924 eV to 1.11 eV, with a bandwidth of 0.186 eV.

The transmission curves under different incident angles were combined into a 2-D contour map to obtain the comparison between the experimental and theoretical results, as shown in figure 7(b). In the 2-D contour map, the brightness of each point denotes transmitted intensity. As can be seen in figure 7(b), the bright part of the map matches the theoretical white line well, indicating that the measured EOTs were obtained from the excitation of the optical MP modes in the diatomic chain of SHRs. The bandwidth of the optical branch can be enlarged if the coupling interaction between elements is increased by changing the length of the slit. In the experiments, another sample with slit length of 50 nm (smaller than the 70 nm slit length of the old sample) was fabricated. The obtained bandwidth was approximately 0.21 eV, which is larger than the bandwidth of the old sample.

The experimental results show that the MP propagation mode in the proposed system can be excited in a broad frequency bandwidth. Figure 7(a) shows the dispersion curves. The dependence of the resonance excitation frequency on the incident angle can be solved numerically by combining Equation (11) and (12), and it shown in two parts because it was divided by the blue light line. The part above the blue line represents the bright MP mode, which can couple to the far-zone optical field. Aside from the EOT reported in this study, the bright MP mode can also be used to produce efficient nanolasers, which have recently aroused research interests [43]. Moreover, the part below the blue line corresponds to the dark MP mode, which cannot be excited by the far-field wave and whose energy does not radiate outwards. The dark MP mode without radiation loss can be greatly amplified using the stimulated emission from an active medium (e.g., quantum dots and the like), similar to how the surface plasmon amplification was achieved using the stimulated emission of radiation (SPASER) achieved in dark SPP mode [44, 45]. This phenomenon can provide a good nanoscale optical source for numerous potential applications in nonlinear optical processes, such as single-molecule detection and florescence imaging.

In the above work, a diatomic chain of SHRs was devised with a unit cell, including two SHRs with equal-length slits and different-sized holes. The MP waves can only be excited through magnetic resonance in the nanoholes, whereas electric resonance does not contribute to excitation. The normal incidence wave cannot be coupled onto MP waves, and the incidence angle should be oblique. In another work [46], a new design for SHR meta-chains was proposed, where the unit cell includes two SHRs with different-length slits and equal-sized holes that are different from our former work [46]. The advantage of the new design is that the coherent MP wave can be excited by both the magnetic resonance in the holes and the electric resonance in the slits. Moreover, the coherent MP in the meta-chains can be excited much more efficiently because of the strong electric resonance in the slits. The excitation can also be realized under normal incidence, and the incidence excitation angle can then be tuned in a wide range from a normal incidence to 40°. In addition, a continuous wide excitation frequency band can be obtained by tuning the incidence angle. The measured dispersion of the coherent MP waves agrees with the calculated theoretical results [46].

2.4 Periodic nanosphere chain on slab

MP resonance can also be established in plasmon molecules created from several coupled nanospheres [47-49]. Such plasmon molecules can be used to form CMROW. In our other recent work [50], a kind of coupled magnetic resonance waveguide is proposed based on a linear chain of contacting nanospheres on a gold slab. Figure 8(a) shows a single unit of the structure with two contacting gold nanospheres placed on a gold slab. The nanosphere had a radius of 200 nm, and the gold layer had a thickness of 50 nm. The nanosphere and the gold layer were separated by a dielectric layer with thickness of 30 nm. A resonance peak, at which the two spheres were shown to exchange current at the contact point, was detected. The excitation also simultaneously induced current on the slab surface. The entire structure can be considered as a closed equivalent LC circuit, as shown in figure 8(b). The two spheres and the slab can be regarded as inductors connected in series, whereas the middle dielectric layer works as a capacitor. However, the resonant current around the closed circuit can induce a strong magnetic field in the area surrounded by the two spheres and the slab, making the structure behave like a magnetic dipole $\vec{m}$. Therefore, this mode was called the MP mode. In the simulations, the relationship between the local magnetic field and the thickness of the dielectric layer is investigated. Under the same incident intensity, the magnetic resonance field decreases with the increase in thickness of the middle layer. Such condition occurs because the EM energy is not contained in the space between the nanospheres and the slab when the gap is increased. Thus, more energy is leaked out and reduces the resonance strength. Once the bottom gold slab is removed, the MP modes become nonexistent because a closed LC circuit cannot be formed without the slab. However, once the dielectric is removed and the nanospheres come into contact with the slab, the MP mode through the LC resonance disappears because of the absence of a capacitor.

The resonances of three nanospheres on a slab were also investigated, as shown in figure 8(c). Given that the former structure, i.e., two nanospheres on a slab, can be considered as a single magnetic resonator, the latter structure, i.e., three nanospheres on a slab, can be perceived as two coupled magnetic resonators. In the simulations, the resonance and field distribution of the latter structure were investigated. The recorded local magnetic field exhibited two resonance peaks. The induced currents in the two LC circuits rotate in the same direction at lower resonance frequencies, enabling the two magnetic dipoles to oscillate in the same phase, as shown in figure 8(d); this mode is called the symmetry mode. In contrast, the induced currents in the two LC circuits rotate in opposite directions at higher frequencies, resulting in the anti-phase oscillation of the two magnetic dipoles, as shown in figure 8(e); this mode is called the anti-symmetry mode.

In our system, the coupling processes between magnetic units include nearest-neighbor exchange current interaction and long-range magnetic field coupling. A semi-analytic model is developed based on the attenuated Lagrangian formalism to provide a good description of the two interactions described above. If L and C are the effective inductance and capacitance of the structure in figure 8(b), respectively, then the Lagrangian formalism of such LC resonator can be expressed as follows: $Lag = \frac{1}{2}L\dot{q}^2 - \frac{1}{2C}q^2$, where q is the oscillating charge in the structure. The structure presented in figure 8(c) can be considered as two connected LC circuits, whose Lagrangian formalism should be expressed as follows:

$$Lag = \frac{1}{2}L(\dot{q}_1^2 + \dot{q}_2^2) - \frac{1}{4C}(q_1^2 + q_2^2) + M\dot{q}_1\dot{q}_2 - \frac{1}{4C}(q_1 - q_2)^2, \tag{13}$$

where $q_1$ and $q_2$ are the oscillating charges in the two LC resonators. The first term represents the kinetic energy in the inductors, and the second term represents the potential electric energy in the gaps under the first and third spheres. The interaction term $M\dot{q}_1\dot{q}_2$ is caused by the MI coupling between the two magnetic resonators. The last term corresponds to the electric potential energy stored in the gap under the second sphere, which can be seen as a shared capacitor of two LC resonators, as shown in figures 8(d) and 8(e). Considering the ohmic dissipation $R = \frac{1}{2}\gamma(\dot{q}_1^2 + \dot{q}_2^2)$ and substituting Equation (13) in the Euler-Lagrange equation yield the following:

$$\frac{d}{dt}\left(\frac{\partial Lag}{\partial \dot{q}_m}\right) - \frac{\partial Lag}{\partial q_m} = -\frac{\partial R}{\partial \dot{q}_m}, \quad (m = 1, 2) \tag{14}$$

A pair of coupled equations can be obtained as follows:

$$\ddot{\mu}_1 + \omega_0^2 \mu_1 + \Gamma \dot{\mu}_1 = \frac{1}{2}\kappa_1 \omega_0^2 (\mu_1 + \mu_2) - \kappa_2 \ddot{\mu}_2$$
$$\ddot{\mu}_2 + \omega_0^2 \mu_2 + \Gamma \dot{\mu}_2 = \frac{1}{2}\kappa_1 \omega_0^2 (\mu_1 + \mu_2) - \kappa_2 \ddot{\mu}_1,$$
(15)

where $\mu_m = A\dot{q}_m$ (m = 1, 2) is the effective magnetic dipole and A is the cross-sectional area surrounded by an induced current in the LC circuit. In Equation (15), $\omega_0^2 = 2/(LC)$ is the eigenfrequency of the single LC circuit and $\Gamma = \gamma/L$ is the damping coefficient caused by ohmic loss. Equation (15) indicates two mechanisms, namely, the exchange of the conduction current coupling and the MI coupling, which are described by two coefficients $\kappa_1$ and $\kappa_2$, respectively. In an ideal circuit, $\kappa_1 = 1/2$ and $\kappa_2 = M/L$ represent the relative strength of the mutual and self inductance of a single unit, respectively. Approximating $\Gamma/2\omega_0 \ll 1$, the two eigenfrequencies can be obtained from Equation (15) as follows: $\omega_1 = \omega_0 \sqrt{(1-\kappa_1)/(1+\kappa_2)}$ and $\omega_2 = \omega_0/\sqrt{(1-\kappa_2)}$. The MP mode at $\omega_1$ is caused by the symmetric resonance of two units with $\mu_1 = \mu_2$, whereas the high-frequency mode $\omega_2$ is caused by the asymmetric $\mu_1 = -\mu_2$. The Lagrangian model above can also be extended to the chain structure shown in figure 9(a). For an infinite chain, let $q_m$ be the oscillation charge in the $m$-th unit. Considering the coupling between magnetic resonators, the Lagrangian formulism can be expressed as follows:

$$\text{Lag} = \sum_m \left[ \frac{1}{2}L\dot{q}_m^2 - \frac{1}{4C}(q_m - q_{m+1})^2 + M\sum_n \frac{1}{n^2}\dot{q}_m \dot{q}_{m+n} \right], \quad (m = 0, \pm 1 \pm 2, \ldots; n = 1, 2, 3\ldots). \quad (16)$$

where the third term indicates the MI coupling between the magnetic dipoles from the nearest neighboring dipoles to the farthest ones. The ohmic dissipation of the whole structure can be expressed as follows:

$$R = \sum_m \frac{1}{2}\gamma \dot{q}_m^2. \quad (17)$$

Substituting Equations (16) and (17) into the Euler-Lagrange equation yields the following:

$$\ddot{\mu}_m + \Gamma\dot{\mu}_m + \omega_0^2 \mu_m = \frac{1}{2}\kappa_1 \omega_0^2 (\mu_{m-1} + 2\mu_m + \mu_{m+1}) - \kappa_2 \sum_n \frac{1}{n^2}(\ddot{\mu}_{m-n} + \ddot{\mu}_{m+n}), \quad (18)$$

where $\Gamma$, $\omega_0^2$, $\mu_m$, and coefficients $\kappa_1$ and $\kappa_2$ are as previously defined. The solutions to Equation (18) have the following form: $\mu_m = \mu_0 \exp(-m\alpha d)\exp(i\omega t - imkd)$, where d is the period of the chain and α is the attenuation per unit length. With $\alpha d \ll 1$ for small damping approximation, the dispersion relationship of the MP mode can be obtained as follows:

$$\omega^2 = \omega_0^2 \frac{1 - \kappa_1[1 + \cos(kd)]}{1 + 2\kappa_2 \sum_n \frac{1}{n^2}\cos(nkd)}, \quad (19)$$

where $\omega_0$ is the eigenfrequency of a single unit. Only the first eight terms of the MI coupling are considered in the succeeding calculations because a larger distance between two dipoles results in a weaker interaction.

A chain of contacting nanospheres that contains 25 linearly arranged gold nanospheres is used in the simulations, as shown in figure 9(a). Excited by a dipole source at a distance of 120 nm in front of the first sphere, the magnetic field at the last nanosphere is recorded, as shown in figure 9(b). The results show that the transmission signal was within the frequency range of 0 THz to 150 THz. The magnetic field along the nanosphere chain at different frequencies can be obtained using the FDTD simulation method. To calculate the dispersion of the MP mode, Fourier transform is used to transform the value of the magnetic field into the wave vector region of the field in ω-k space [51]. The Fourier transform can be expressed as follows:

$$H(\omega, k) = \int H(\omega, x) e^{ikx} dx. \qquad (20)$$

The Fourier transform is processed along the chain and yields the dispersion relation. The results are shown as a grey map in figure 9(c). The dispersion of the MP mode is very similar to that of a surface plasmon, in which the wave vector k increases with ω from 0 THz to 150 THz. The theoretical dispersion result based on Equation (19) is also deduced, as represented by dots in figure 9(c). The Lagrangian model agrees with the simulated results quite well. The Lagrangian model used in this study can be generalized to include other possible coupling interactions, such as plasmon-mechanical or plasmon-acoustical effects, in future coupled systems.

Based on the dispersion relation of the MP mode in Equation (19), the group velocity can be calculated as follows:

$$V_g = \frac{\partial \omega}{\partial k} = \frac{\omega_0^2 d}{2\omega} \cdot \frac{\kappa_1 \sin(kd) \cdot \left[1 + 2\kappa_2 \sum_{n=1}^{\infty} \frac{1}{n^2} \cos(nkd)\right] + 2\kappa_2 \left[1 - \kappa_1(1 + \cos(kd))\right] \cdot \sum_{n=1}^{\infty} \frac{1}{n} \sin(nkd)}{\left[1 + 2\kappa_2 \sum_{n=1}^{\infty} \frac{1}{n^2} \cos(nkd)\right]^2}. \qquad (21)$$

where only the eight nearest dipole coupling interactions are considered. Figure 9(d) shows the calculated dispersion property of group velocity. The results show that the group velocity was very small ($V_g = 0.1c$) at approximately $k = \pi/d$ ($\omega = 140\,\text{THz}$). The very small group velocity of the MP mode can be obtained from the designed structures. The slow light effect has been reported in various physical systems, including atomic gases, optical fibers, photonic crystals, and plasmon systems. In the current study, the proposed structure also demonstrates the dispersive slow wave effect in the subwavelength scale via MP excitation. Although the spin waves in magnetic materials have many interesting properties in the microwave range, the analog of spin waves in the infrared or THz region proves to be an interesting topic and may exhibit new properties. In this study, the slow wave is caused by the coupling effect between magnetic resonators. This wave mimics the slow spin waves in the infrared or THz region that does not occur naturally. Furthermore, given that the magnetic resonator is completely designed artificially and that the coupling interaction can be tuned at will, the dispersion of the slow wave effect can be controlled completely by altering the structural parameters. Then, a slow spin wave can be obtained at the infrared or THz region.

## 2.5 Periodic nanosandwich chain

The nanosandwich structure, as one of the basic building blocks in plasmonics, is recommended to be used in making a subwavelength waveguide in the high-frequency regime because of its simple structure and high working frequency regime. Figure 10(a) shows the geometry of a single nanosandwich that is composed of two metallic nanodiscs and a dielectric middle layer [51]. The anti-parallel currents in the metallic slabs induce a high intensity and confined magnetic field at a certain frequency, which can be seen as a magnetic atom. Figures 10(b-d) respectively show the frequency spectrum and field distribution of such a nanosandwich. Such a magnetic atom can be used to construct a linear magnetic chain. An MP propagation mode is established in the 1D system because of the near-field electric and magnetic coupling interactions. A strong local magnetic field can be obtained in the middle layer at a specific frequency when it is excited by an EM wave, as shown in figure 10(e). Figure 10(f) shows the corresponding electric fields for such magnetic plasmon resonance mode. It should be noted that such an MP waveguide is a subwavelength, the energy flow cross-section of which is plotted in figure 10(g). The field is confined in a small area smaller than the wavelength scale. The wave vectors of the MP waveguide at different EM wave frequencies can be calculated using a Fourier transform method to obtain the dispersion property of the MP wave, as shown by the white line in figure 10(h). The light line in free space is represented by the black dotted line in figure 10(h). The light line divided the MP curve into two parts. The part above the light line corresponds to the bright MP modes whose energy radiated out from the chain, whereas the part below the light line corresponds to the dark MP modes whose energy can be confined within the chain. The bright MP modes were much weaker than the dark MP modes in terms of their leaky property. Therefore, only the EM waves in the frequency range of the dark MP modes can be transferred efficiently without radiation loss.

## 3. Aperiodic CMROW

In graded waveguides and metamaterials, we can control the effective index continuously. Through this method, we can slow down the speed of light and trapping the light in the structures. The graded system can be used to photon storage and nonlinear optical processes [52]. In [53], we designed a graded nanosandwich waveguide. Once the results for a mono-periodic chain of nanosandwiches have been generalized to graded structures, some new interesting properties, such as slow group velocity and a new type of field distribution, can be obtained in more complex structures. Then, the chain composed of such nanosandwiches with the spacing between nanosandwiches being linearly increasing along the chain, which indicates a graded changing coupling between nanosandwiches, can be investigated. The spacing $d_m$ obeys the following rule: $d_m = 225 + 100\times(m-1)$, where $m$ denotes the spacing between the $m$-th and ($m$+1)-th nanosandwich. Figure 11(a) shows the geometry of the chain with 41 nanosandwiches, and figure 11(b) shows the dispersion relation of the graded chain. The MP modes can be divided into three parts, namely, gradon (the special mode that belongs to the graded structure), extended mode, and evanescent mode. Figure 11 (b) also show the different propagation distance for these three modes, in which the distance is denoted by the number of periods along the chain. The field distributions of the three parts of the MP modes are quite different from one another, and the location of the field of the gradon is strongly dependent on the frequency, as shown in figure 11(c). Above the light line at 248 THz, the MP mode is an evanescent mode, with the field amplitude decreasing exponentially. At 266 THz, the MP mode is an extended mode; the field can propagate throughout the chain. At 280 THz, although the MP mode is below the light line, the field in the chain cannot reach the end of the chain but stops at the middle of the waveguide, which is a typical field localization in the graded structure. Since this mode is at the high frequency region of the MP mode band, it is called "light gradon." A wavelength selective switch can be managed by employing this property. Three-and four-port switches can be realized in the graded nanosandwich chain. Figure 11(c) shows the field distributions of the magnetic field corresponding to different modes of the switches. Some new interesting properties, such as slow group velocity and band folding of MP waves, can be obtained in such complex structures.

## 4. Nonlinear CMROW

The loss that includes the large scattering loss introduced by the micro-fabrication and the ohmic loss of the metal component, especially at the light frequency region, prevents the subwavelength plasmonic waveguides from being realistically applied. Usually, for a plasmonic waveguide, the propagation length is less than 50 micrometers. The combination of metallic structures with gain materials is a promising method for compensating the loss in plasmonic systems [45, 54-56]. In our recent work [57], a magnetic plasmon nanolaser is reported based on double resonance nanosandwich structures. In another work of ours [58], the compensation effect in an MP waveguide combined with the ytterbium-erbium codoped gain material, Er:Yb:YCOB, in which the lasing case is found, is investigated.

Figure 12(a) shows the geometry of the subwavelength MP waveguide. The nanosandwich is composed of two metallic rectangular slabs. The middle layer and the surrounding environment are both chosen to comprise the ytterbium-erbium codoped gain material, Er:Yb:YCOB, with a refractive index of 1.3. The gain waveguide system is placed on the $SiO_2$ substrate with a refractive index of 1.5. In such a nanosandwich waveguide, the collective magnetic resonance, MP mode, can be excited using a near-field source placed at the input of the waveguide. Figure 12(b) shows the energy density distribution of such MP mode with a wavelength of 1550 nm. The nanosandwich waveguide can also sustain high-order modes. Figure 12(c) shows the energy density distribution of the high-order mode of the waveguide with a wavelength of 980 nm, which is exited by a plane wave source incident on the entire waveguide plotted in figure 12(a). Since the nanosandwich waveguide can be considered as a chain of coupled resonators, as shown in figure 12(d), the energy propagation along the waveguide can be described as follows:

$$\frac{\partial N^i}{\partial t} = \frac{N^{i-1}}{\tau_{Prop}} - 2\frac{N^i}{\tau_{Prop}} + \frac{N^{i+1}}{\tau_{Prop}} - \frac{N^i}{\tau_{Loss}} = 0 \qquad (22)$$

where $N^i$ denotes the number of photons of the signal in the $i$-th nanosandwich and $\tau_{Prop}$ and $\tau_{Loss}$ correspond to the propagating and loss processes, respectively. In the steady-state case, Equation (22) is zero.

Equation (22) will change and the term of gain effect should be added into it when the waveguide is combined with the gain material. In steady-state conditions, by neglecting the populations in levels $^4I_{11/2}$, $^4I_{9/2}$, and $^4F_{9/2}$ and corresponding back-transfer processes because of the fast non-radiative decay in such levels, the simplified rate equations can be expressed as follows [59-61]:

$$\frac{\partial N_{2Y}}{\partial t} = \sigma_Y v_p F_p N_p f_p (N_{1Y} - N_{2Y}) - k_1 N_{2Y} N_{1E} - k_2 N_{2Y} N_{2E} - \frac{N_{2Y}}{\tau_{2Y}} = 0$$

$$\frac{\partial N_{2E}}{\partial t} = k_1 N_{2Y} N_{1E} - \sigma_E v_s F_s N_s f_s (N_{2E} - N_{1E}) - \frac{N_{2E}}{\tau_{2E}} - 2CN_{2E}^2 = 0 \quad (23)$$

$$\frac{\partial N_s}{\partial t} = \sigma_E v_s F_s \int (N_{2E} - N_{1E}) N_s f_s dV - \frac{N_s}{\tau} = 0$$

where $N_{ix}$ and $\tau_{ix}$ represent the population density and lifetime of the corresponding levels of Er and Yb (given in the figure 13 (d)). $\tau$ is the decay time of the MP waveguide mode in the chain. $k_1 = k_2 = 5.0 \times 10^{-21}$ $m^3/s$ are the coefficients of the two energy transfer processes. $C$ is the up-conversion rate and is equal to $1.3 \times 10^{-23}$ $m^3/s$. $v_p$, $N_p$, and $f_p$ represent the group velocity, total photon number, and normalized spatial intensity distributions of the pump light (980 nm). $v_s$, $N_s$, and $f_s$ represent the corresponding parameters of the signal light (1550 nm). $f_p$ and $f_s$ are normalized as $\int f_p dV = 1$ and $\int f_s dV = 1$, respectively, where $V$ is the volume. In addition, in steady-state conditions, the approximate expressions $N_{1E} + N_{2E} \approx N_E$ and $N_{1Y} \approx N_Y$ can be provided. In the calculations, the values of $\sigma_E$, $\sigma_Y$, $\tau_{2E}$, and $\tau_{2Y}$ were fixed at $5.0 \times 10^{-25} m^2$, $8.0 \times 10^{-25} m^2$, $5.0 \times 10^{-3} s$, and $2.6 \times 10^{-3} s$, respectively [57, 59-62]. Here, $F_p$ and $F_s$ are the Purcell factors for the pump and signal respectively, which can be calculated as $F_p = 3Q_p \lambda_p^3 / (4\pi^2 V_{pm} n^3)$ and $F_s = 3Q_s \lambda_s^3 / (4\pi^2 V_{sm} n^3)$ [63, 64], where $n$ is the refractive index of the gain material and $\lambda$ is the wavelength. The quality factor $Q_p$ ($Q_s$) and the effective mode volume of laser mode $V_{pm}$ ($V_{sm}$) are determined by the decay time of the mode and the field confinement, respectively. Both $Q_p$ ($Q_s$) and $V_{pm}$ ($V_{sm}$) can be calculated in the simulations. In the simulations, the coefficient $3\lambda^3 / (4\pi^2 V_{pm} n^3)$ and $3\lambda^3 / (4\pi^2 V_{sm} n^3)$ for signal and pump light were 1/40 and 1/20, respectively. The group velocities of pump light and signal light were also calculated to be $1.0 \times 10^8 m/s$ and $0.5 \times 10^8 m/s$, respectively.

Considering the gain effect, the propagation equation can be modified as follows:

$$\frac{\partial N_s^i}{\partial t} = \frac{N_s^{i-1}}{\tau_{Prop}} - 2\frac{N_s^i}{\tau_{Prop}} + \frac{N_s^{i+1}}{\tau_{Prop}} - \frac{N_s^i}{\tau_{Loss}} + \sigma_E v_s F_s \int (N_{2E} - N_{1E}) N_s^i f_s dV = 0 \quad (24)$$

Equation (24) is zero in the steady-state case. In this study, the MP mode was chosen as the signal light and the high-order mode as the pump light, leading to the larger efficiencies of pumping and radiation [57]. In general, the $Yb^{3+}$ concentration is an order of magnitude higher than the $Er^{3+}$ concentration. In the calculations, the $Yb^{3+}$ concentration was fixed at $5.0 \times 10^{27}$ ions/$m^3$ [57], and the pumping power on a single nanosandwich was fixed at 0.05 mW. Different $Er^{3+}$ concentrations impose different compensation effects against the loss in waveguide. Figure 13(a) shows the normalized number of photons in nanosandwiches along the waveguide with different $Er^{3+}$ concentrations. A larger $Er^{3+}$ concentration leads to higher compensation. The propagation length doubled when $N_E$ was increased to $3.0 \times 10^{26}$ ions/$m^3$ with respect to the case of $N_E = 0$, as shown in figure 13(a). Moreover, we can see from the same figure that increasing the doping concentration above

$N_E$=3.0×10$^{26}$ ions/m$^3$, we can seriously increase the propagation length, such as in the case of $N_E$=3.5×10$^{26}$ ions/m$^3$. Although the energy of signal decreases along the waveguide, it does not become zero but rather stops at a certain value. This phenomenon is attributed to the saturation of stimulated emission radiation under certain concentration of gain ions and pumping power that can only afford a low signal. On the other hand, the fabrication scattering and nanosized metallic structure loss were considered by reducing the decay time of the MP mode. The energy that transports along the waveguide with different decay times of MP mode from 100 fs to 40 fs, which is the typical range in metamaterial and plasmonic structures, was also considered. A longer propagation length can be obtained with longer decay time of MP mode, as shown in figure 13(b). Lower loss clearly leads to more evident compensation effect. Moreover, the saturation phenomenon was obtained with longer decay time of $\tau$ = 100 fs, as shown in figure 13(b).

The loss is largely compensated by the gain effect, especially in the case of high Er$^{3+}$ concentration of $N_E$ and long decay time of $\tau$. In fact, the Er$^{3+}$ concentration can be further increased to approximately 10$^{27}$ icon/m$^3$ [65], and the decay time of the MP mode (signal) calculated directly from the simulations was larger than what was chosen in figure 13 (approximately 110 fs). Therefore, the gain effect defeats the loss effect and leads to the stimulated emission. In other words, the signal is enhanced as it propagates along the waveguide, similar to that of a fiber amplifier. The loss in waveguide can be largely compensated by tuning the doping concentration of Er$^{3+}$ and decay time of the signal. The compensation cases can be divided into two types because of the saturation effect. Moreover, the gain effect can overcome the loss in the waveguide when the parameters exceed a certain threshold, leading to the amplification of signals along the waveguide similar to that of the fiber amplifier. This property has potential application in plasmonics-integrated optical circuits and metamaterials.

## 5. Quantum CMROW

Since the first demonstration of the plasmon-assisted entangled photons in perforated metal film, the quantum characteristics of plasmonic system and metamaterials have been continuously reported and investigated for their potential applications in quantum information techniques [45, 66-74]. The quantum generator made from the surface plasmon amplification by stimulated emission of radiation (SPASER) system has recently been introduced, in which a generalized quantum treatment of surface plasmon was introduced using the spectral representation method [66, 67]. All the improvements above require a profound understanding of the fundamental quantum properties of coupled metamaterials. Therefore, the interaction between coupled metamaterials and other materials should also be investigated further.

In our recent paper [75], the interaction between quantum dots and a 1D coupled metamaterial composed of a chain of nanosandwiches was investigated. Figure 10(a) shows the geometry of a single nanosandwich. The nanosandwiches are closely placed in a line to form a 1D CMROW, and the nanosandwich is placed on a silica substrate. When the middle layer of the nanosandwich is filled with a non-metallic material, the magnetic resonance can be formed by the excited magnetic loop composed of currents in two separate metal layers and the displacement currents in the outside surrounding [51]. The electromagnetic field is highly confined in the middle layer, which is filled with the active material, semiconductor PbS quantum dot material, with an emission wavelength of approximately 1550 nm, telecom wavelength, and electric permittivity of $\varepsilon_{PbS} = 23$. The quantum dots are densely packed in vacuum in the middle layer [66, 67]. In the coupled metamaterial, the magnetic resonances in nanosandwiches couple with each other and form a coupled magnetic plasmons (CMP), which is a collective magnetic resonance throughout the chain [51]. This study investigates two cases of the coupled metamaterials, namely, the 1D coupled metamaterial embedded in vacuum (Case I) and the one embedded in quantum dot material (Case II).

A full quantum treatment based on the quantization on the Hamiltonian of the CMP, which can be considered as a kind of excitation in the artificial material [73, 74, 76], is used to investigate the interaction between CMROW and the quantum dots. The Lagrangian formalism of the nanosandwich chain can be expressed as follows:

$$\text{Lag} = \sum_m \left[ \frac{L}{2}\dot{Q}_m^2 - \frac{1}{2C}Q_m^2 + \frac{M_h}{2}(\dot{Q}_m\dot{Q}_{m+1} + \dot{Q}_m\dot{Q}_{m-1}) - \frac{M_e}{2}(Q_mQ_{m+1} + Q_mQ_{m-1}) \right] \quad (25)$$

Here, $M_h$ and $M_e$ describe the magnetic and electric coupling between the unit cells, respectively. Using a Legendre transformation, $\text{Ham} = \sum_m P_m\dot{Q}_m - \text{Lag}$, the Hamiltonian of the coupled metamaterial that plays a much more important role than the Lagrangian formulism in solid-state physics can be obtained. Using the generalized momentum $P_m = \partial \text{Lag}/\partial \dot{Q}_m$ correlated with $Q_m$, the Hamiltonian can be expressed as follows:

$$\text{Ham}(Q_m,\dot{Q}_m) = \sum_m \left[ \frac{L}{2}\dot{Q}_m^2 + \frac{1}{2C}Q_m^2 + \frac{M_h}{2}(\dot{Q}_m\dot{Q}_{m+1} + \dot{Q}_m\dot{Q}_{m-1}) + \frac{M_e}{2}(Q_mQ_{m+1} + Q_mQ_{m-1}) \right] \quad (26)$$

A Fourier transformation was applied to Equation (26) to obtain the following expression:

$$\text{Ham}(Q_k,\dot{Q}_{-k}) = \sum_k \left[ \frac{L}{2}\dot{Q}_k\dot{Q}_{-k} + M_h\cos(kd)\dot{Q}_k\dot{Q}_{-k} + \frac{1}{2C}Q_kQ_{-k} + M_e\cos(kd)Q_kQ_{-k} \right] \quad (27)$$

where $d$ refers to the period of the metamaterial and the Fourier expansion $Q_m = \frac{1}{\sqrt{M}}\sum_k Q_k e^{ikR_m}$ is used (M is the number of the resonators). Charge $Q_k$ has a canonically conjugate variable $P_k = \partial \text{Lag}/\partial \dot{Q}_k = \left(\frac{L}{2} + M_h\cos(kd)\right)\dot{Q}_{-k}$. Considering the quantum condition, $\hat{Q}_m$ and $\hat{P}_m$ possess the commutation relation $[\hat{Q}_m,\hat{P}_m] = i\hbar$ [77, 78]. The commutator between $\hat{Q}_k$ and $\hat{P}_k$ was derived to be $[\hat{Q}_k,\hat{P}_k] = i\hbar$. In the derivation, the unitary condition $\frac{1}{M}\sum_m e^{i(k+k')md} = \delta_{k,-k'}$ was used. A Bogoliubov transformation was performed to the Hamiltonian in Equation (27) by introducing a set of creation and annihilation operators, namely, $\hat{a}_k = U_k\hat{Q}_k + iV_k\hat{P}_{-k}$ and $\hat{a}_k^+ = U_k\hat{Q}_{-k} - iV_k\hat{P}_k$, with parameters $U_k = (\hbar)^{-1/2}\sqrt{\xi}$, $V_k = (\hbar)^{-1/2}/\sqrt{\xi}$, and $\xi = \sqrt{[\frac{1}{2C} + M_e\cos(kd)][\frac{L}{2} + M_h\cos(kd)]}$. Finally, the Hamiltonian of a coupled metamaterial in number representation can be obtained as follows:

$$\text{Ham} = \sum_k \left(\hat{a}_k^+ \hat{a}_k + \frac{1}{2}\right)\hbar\omega_k \tag{28}$$

Figure 14 shows the derivation process from Lagrangian model to Hamiltonian model, which was used to make the mathematical formalism above more understandable. The quantum description of the excitation in a coupled metamaterial can be obtained using Equation (28). For convenience, the concept of 'quasi-particle' can be used to provide an institutive picture of the quantum property of the excitation, CMROW, in such 'meta-solid.' $\hat{a}_k^+$ and $\hat{a}_k$ are the creating and annihilating operators that indicate the creation and destruction of a 'quasi-particle,' respectively, with momentum $\hbar k$ in the coupled metamaterial and number operator of $\hat{a}_k^+\hat{a}_k$. The 'quasi-particle' describes the collective resonance behavior of the CMROW throughout the entire solid-state-like metamaterial when the coupling between unit cells exists. Once the coupling shrinks to zero, the metamaterial returns to the free-gas case, in which the model will simply correspond to the excitation of the unit cell itself. Such quantum treatment can also be used in plasmonic structures created from metallic nanostructures. A possible experimental proof of the quantum characteristic of metamaterials can be obtained by measuring the second-order quantum coherence function $g^{(2)}(0)$ in an attenuated-reflection set-up. For a quantum state, $|n\rangle$, $g^{(2)}(0) = 1 - 1/n < 1$ indicates a quantum property that can be measured directly in a practical experiment [73, 74].

The interaction Hamiltonian of the active system can be obtained after introducing the Hamiltonian and the full-quantum treatment of the coupled metamaterial. The interaction Hamiltonian of the system is expressed as follows: $\text{Ham}_{int} = \sum_r \mathbf{E} \cdot \mathbf{d}$, where $E$ denotes the electric field in metamaterial, $d$ refers to the dipole moment of excitation in the quantum dot, and the summation corresponds to all quantum dots in the system [70-72]. After some derivations, the quantized interaction Hamiltonian can be obtained as follows:

$$\text{Ham}_{int} = \hbar \sum_k [G_k(\hat{a}_k^+ \hat{\sigma}_k^- + \hat{a}_k \hat{\sigma}_k^+)] \tag{29}$$

where the coupling constant $\mathbf{G}_k$ is equal to $\sqrt{\int (\rho_2(\alpha) - \rho_1(\alpha))(\varphi_k(\alpha) \cdot \mathbf{d}_{1,2})^2 d\alpha^3} / \hbar$, which is of crucial importance in describing the strength of the interaction between the metamaterial and the quantum dots. $\varphi_k(\alpha)$ is the eigenstate of the excitation corresponding to the electric field distribution with the energy normalized to $\hbar\omega_k/2$, with $\alpha$ being the position of the quantum dot in the unit cell. $\rho_2$ and $\rho_1$ represent the population densities of the two levels. The transition operators $\hat{\sigma}_k^+$ and $\hat{\sigma}_k^-$ indicate the creating and annihilating operators of the quantum dots that belongs to the whole system with momentum $\hbar k$, respectively. The rotating wave approximation was used to eliminate the non-conserving energy terms. Considering the maximum population inversion yields $\rho_2 - \rho_1 \approx \rho$, where $\rho = \rho_2 + \rho_1$ is proportional to $r^3$ and a moderate choice on the radius of the quantum dot was taken as $r \approx 2.5\ nm$. The dipole moment was chosen to be $|\mathbf{d}| = 1.9 \times 10^{-17}\ \mathbf{esu}$ and the coupling constant was calculated.

The coupled kinetic equation for the emission processes of quasi-particles of CMROW can be obtained from the interaction of the Hamiltonian and the coupling constant **G**$_k$. Under strong optical-pumped or electric-pumped conditions, the energy absorbed by the quantum dots is very large and saturated and the level of excitation is quite large. The number of excitation can be assumed to remain constant at $\dot{\sigma}_k^z \approx 0$ and only consider the change in the quasi-particles of the coupled metamaterial. In addition, the homogeneous broadening of the quantum dot spectrum is much narrower than the excitation in the coupled metamaterial, and thus, it is considered as a continuous radiation field and a narrow quantum-dot spectral case. Therefore, the stimulated emission rate must be integrated in a narrow frequency range, and the density of the state must be considered. Finally, a stimulated emission rate B was obtained as $2|G_k|^2 Md/v_g$, where $M$ is the total number of unit cells and $v_g$ is the group velocity derived from the dispersion relation obtained above. This results is consistent with the results of Fermi's golden rule. Moreover, the damping term $\kappa$ can be considered as $\kappa = \gamma_{meta}$, where $\gamma_{meta} = 1/\tau_{meta}$ represents the decay rate. Figures 15(a) and 14(b) show the stimulated emission coefficient and lifetime of quasi-particle of coupled metamaterial in both systems with different spacings. The coupling between the unit cells as well as the lifetime of the cells decreased with increasing spacing. As can be seen in figure 15, the stimulated emission coefficients in both cases decreased with increasing spacing. The interaction in Case II became stronger than that in Case I with increasing number of available quantum dots, leading to a larger B. The gain of the system is defined as $\Gamma = B/\kappa - 1$, and thus, the cases with $\Gamma > 0$ correspond to the amplification condition. Furthermore, the stimulated emission coefficient B is dependent on the number of unit cells according to its expression. Thus, the stimulated emission can be increased further by increasing $M$. When $M$ is large, $\Gamma > 0$ can easily be obtained and the amplification by stimulated emission radiation occurs.

## 6. Outlook

Some important progress on CMROW have recently been reported. The Halas group reported a metallic disc array to transport magnetic plasmon mode [79], and they were able to obtain Y-splitter and MZ interference devices by using such system. CMROW can be applied to many other nanocircuits [80]. In the current study, only the CMROW in planar substrate was considered, although coupling can also happen between magnetic resonators in a 3D configuration [39, 81, 82]. CMROW can be fabricated in a 3D configuration using these techniques in the future. Almost all 3D plasmonic structures can be seen as stereometamaterials that possess many similar optical properties as stereochemistry with properties that are mainly determined by the 3D configuration instead of their elements. CMROWs can be fabricated to mimic double helix DNA structures or other complex polymer structures. Another important progress on quantum CMROW is a reported experiment work about quantum interference in CMROW, which shows the possibility of application of CMROW in quantum optics [83].

CMROWs will continue to face many challenges in the future. The most important challenge is the ohmic loss caused by metallic materials during light propagation. Ohmic loss is always serious in the visible and infrared ranges. The ohmic loss will be much smaller in the microwave and THz ranges. Moreover, the geometry size of magnetic resonator can be quite large. CMROW structures can easily be obtained using some simple and commercial fabrication techniques. Most applications of CMROW should be performed in the THz and microwave ranges. In contrast to traditional transmission lines, CMROW is a kind of periodic system. The dispersion of waveguide mode and the light velocity can be tuned by changing the coupling between magnetic resonators, providing CMROW a sufficient space to obtain various controllable devices.

**Acknowledgement.** This work was financially supported by the National Natural Science Foundation of China (No. 11021403, 11074119 and 60990320), the National Key Projects for Basic Researches of China (No. 2010CB630703, 2012CB921500 and 2012CB933501) and the Program for New Century Excellent Talents of China.

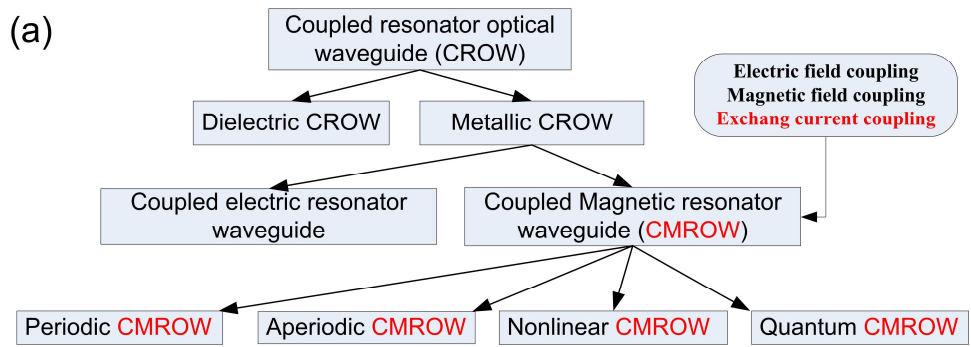

**Figure 1** Several kinds of coupled magnetic resonator waveguides

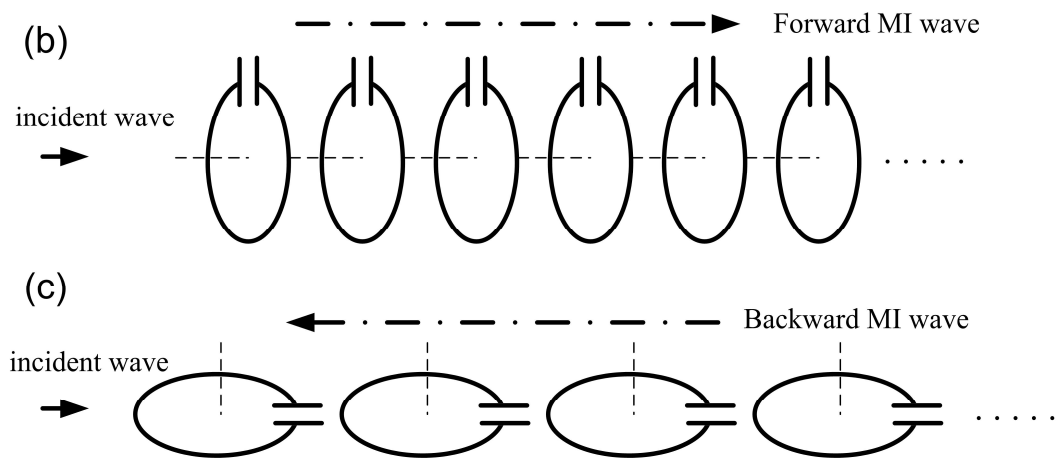

**Figure 2** (a) Forward MI waves on an axial array of SRR chains; (b) backward MI wave in a planar of SRR chains.

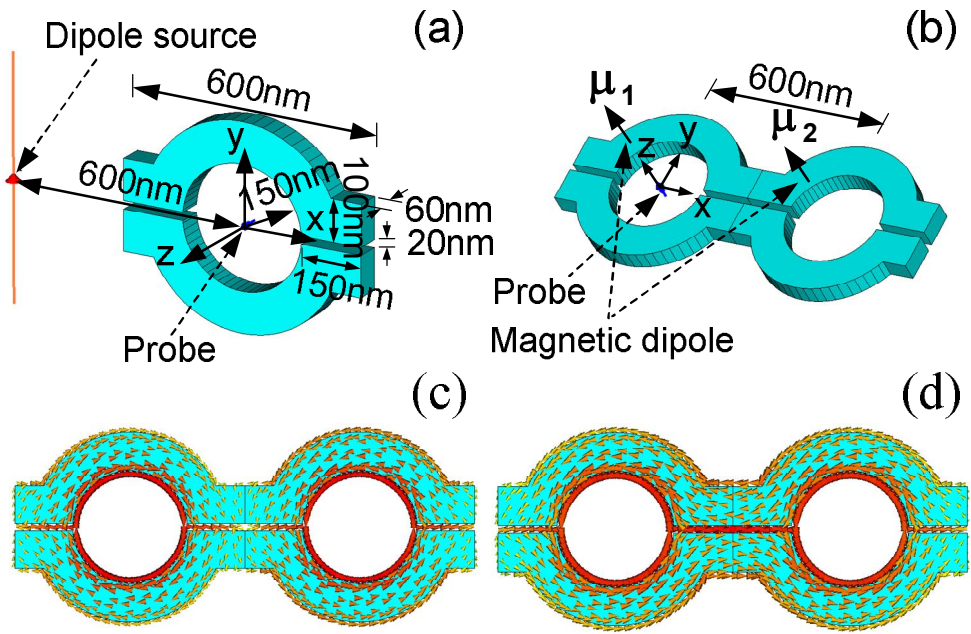

**Figure 3** (a) Single SRR; (b) two connected SRRs; (c) anti-symmetry mode; (d) symmetry mode. From [36].

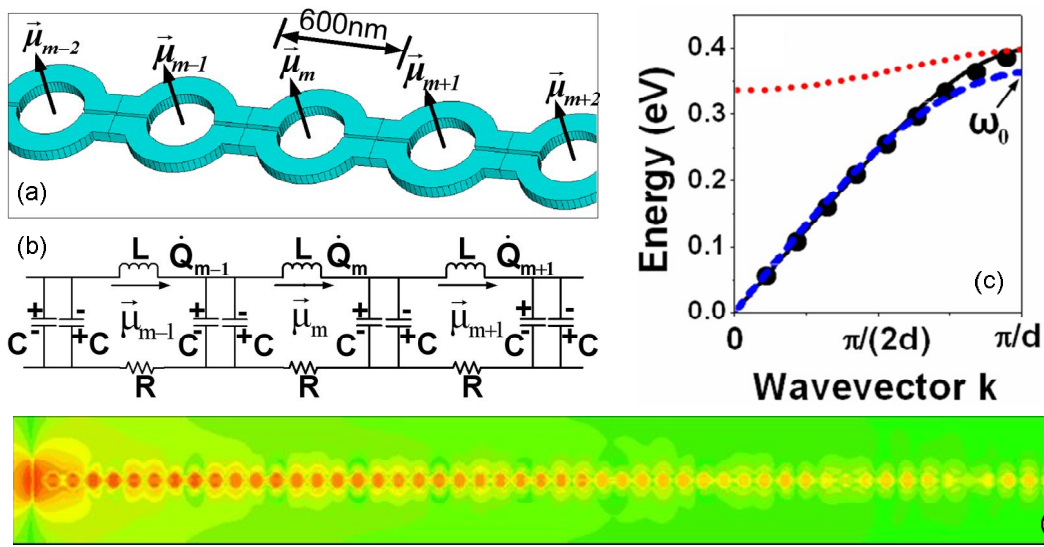

**Figure 4** (a) Structure of connected SRR chain; (b) equivalent LC circuit of SRR chain; (c) dispersion curve of magnetic plasmon. The analytical results, including conduction current and magneto-inductive interactions and the solid curve match well with the FDTD numerical data (circles). The predicted MP characteristics was based singularly on exchange current interactions ($\kappa_2 = 0$), and the magneto-inductive interactions ($\kappa_1 = 0$) are presented with dashed and dotted curves, respectively. (d) magnetic field profile of a magnetic plasmon mode. From [36].

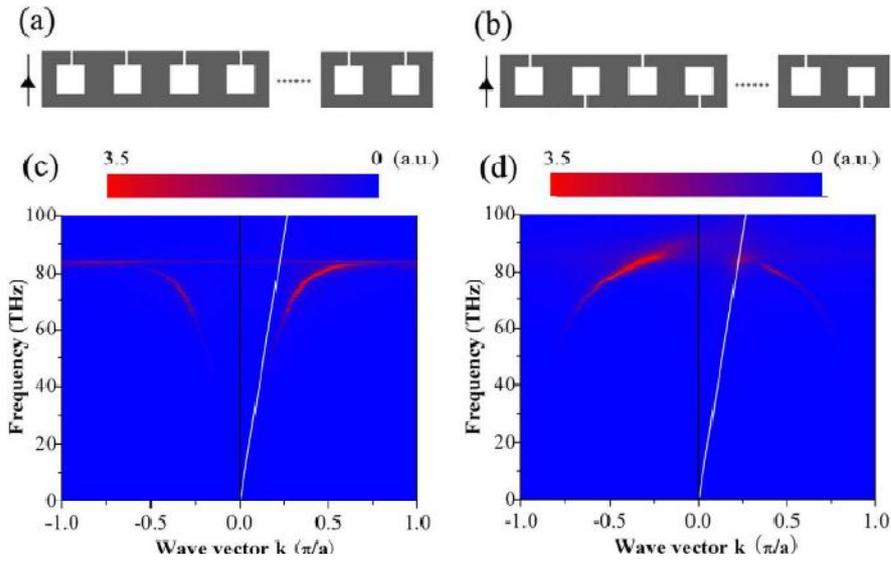

**Figure 5** Subwavelength waveguides constituted by the SRR chains with (a) homo-connection and (b) anti-connection. (c) and (d) are the Fourier transformation map in the ω-$k$ space corresponding to the waveguides in (a) and (b), respectively. From [40].

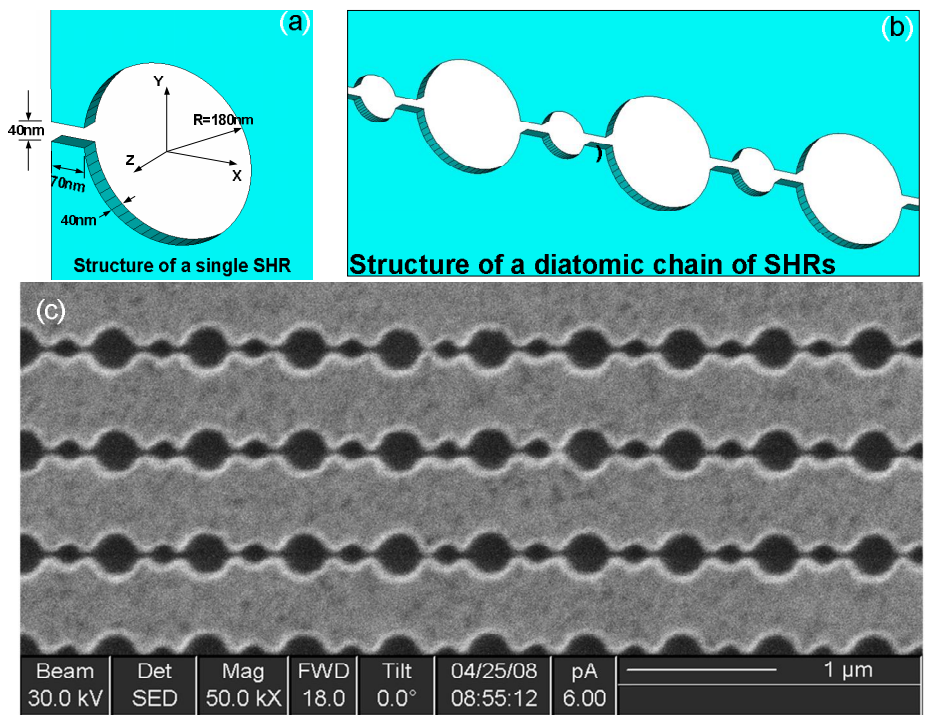

**Figure 6** (a) Structure of a single SHR; (b) structure of a diatomic chain of SHRs; (3) FIB image of the fabricated SHR chain. From [41].

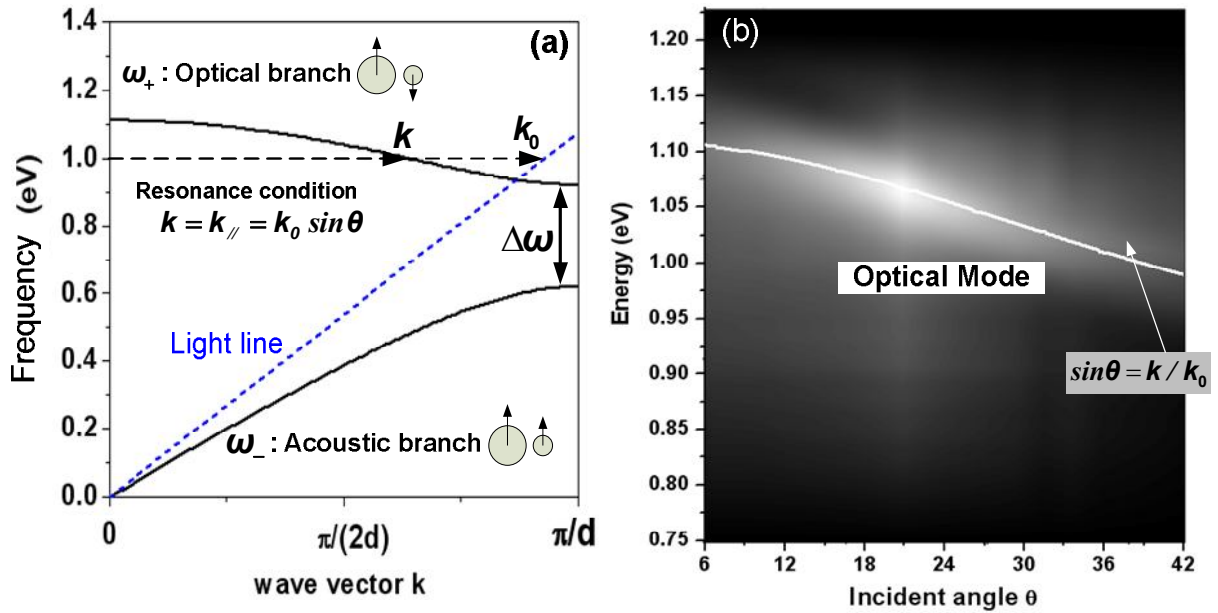

**Figure 7** (a) Dispersion curves for the MP modes in the diatomic chain of SHRs; (b) measured transmission map and the calculated angular dependence curve of the optical MP mode. From [41].

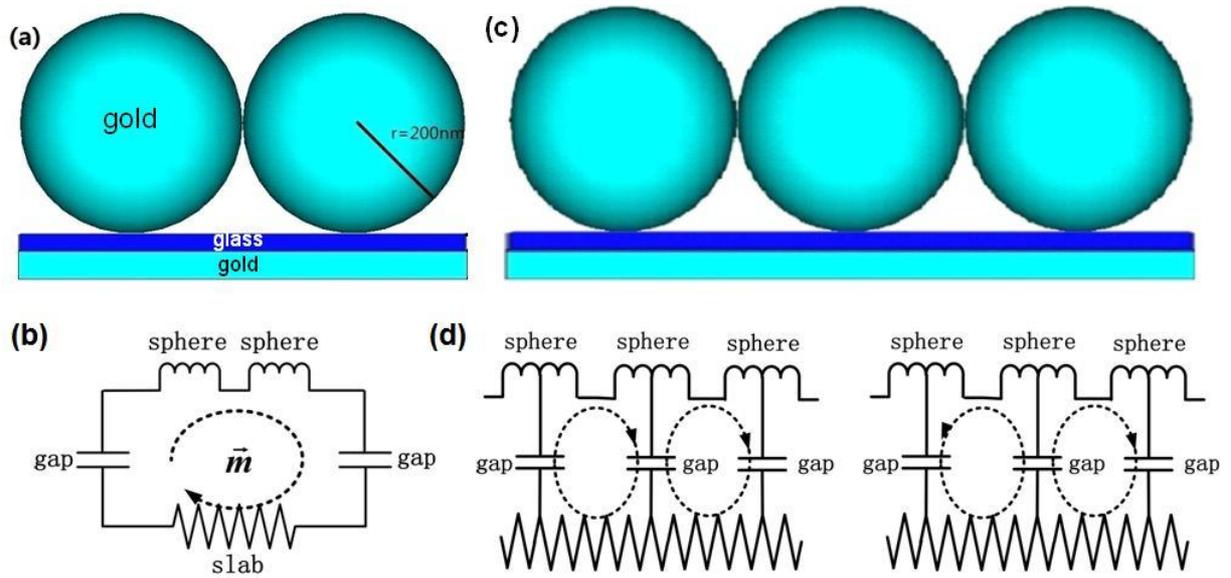

**Figure 8** (a) Two contacting gold spheres on glad slab; (b) equivalent circuit of the structure given in (a); (c) three contacting spheres on the slab; (d) equivalent circuit of symmetry mode and (e) anti-symmetric mode. From [50].

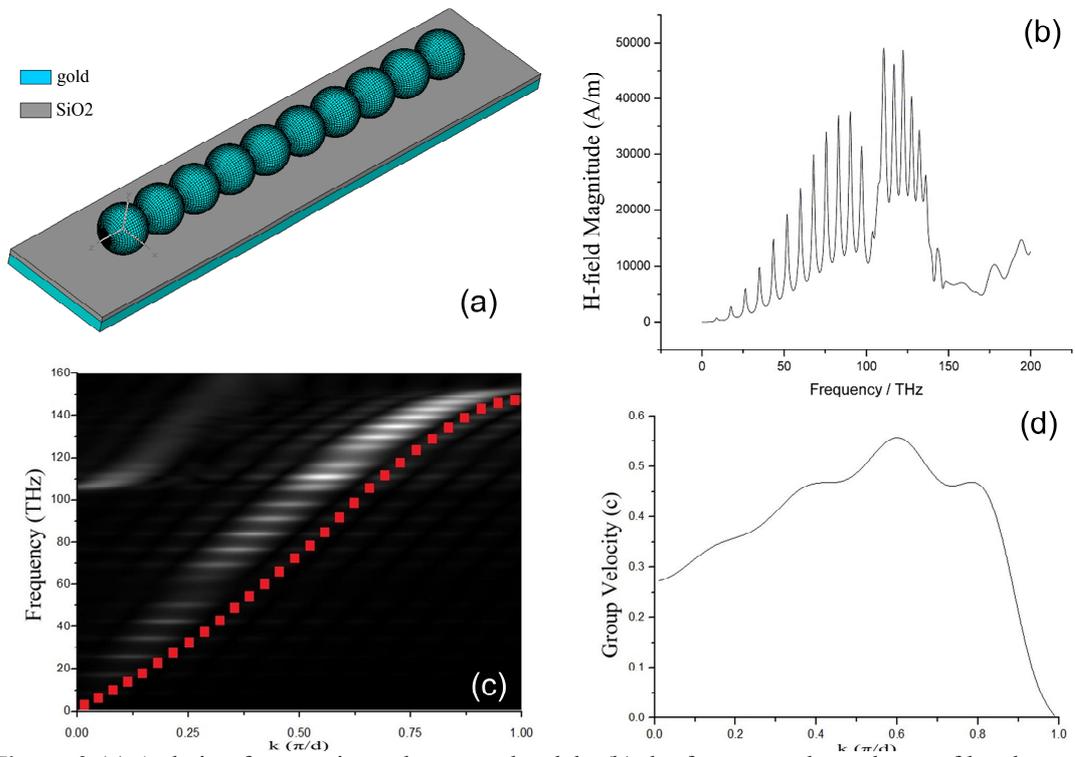

**Figure 9** (a) A chain of contacting spheres on the slab; (b) the frequency dependence of local magnetic field at the end of the chain (recorded using a probe at the last gold sphere); (b) the dispersion curve of coherent magnetic plasmon modes (grey map: simulated result; red square-dot line: calculated results based on Lagrange mode); (d) dispersion of group velocity. From [50].

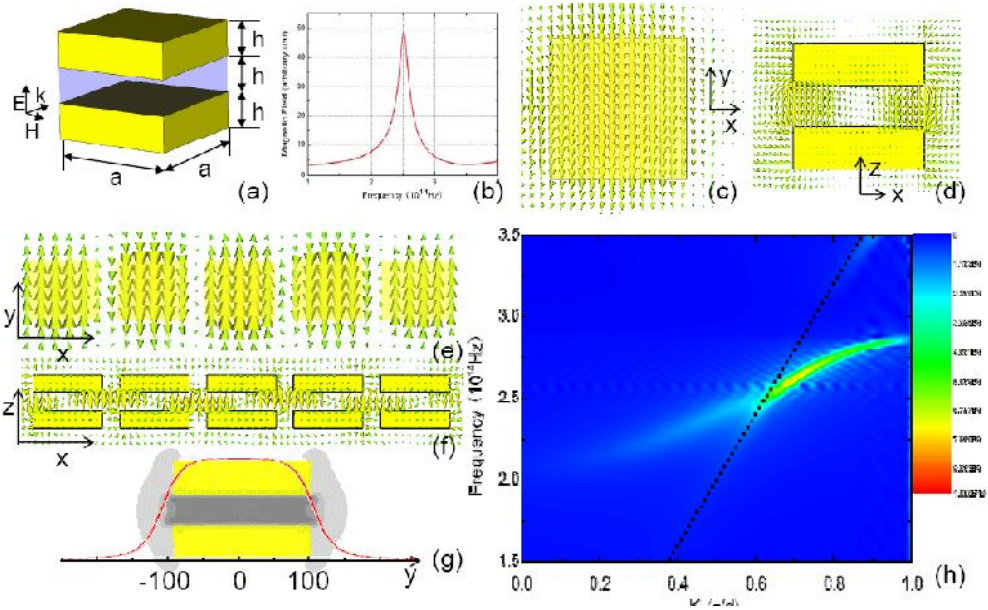

**Figure 10** (a) The geometry of a nanosandwich; (b) the spectrum of magnetic field in the nanosandwich; (c) and (d) show the magnetic field and electric field distribution at the magnetic resonant frequency, respectively; (e) and (f) represent the field distribution of a nanosandwich chain; (g) the energy flow cross-section; (h) the dispersion relation of the nanosandwich chain. From [51].

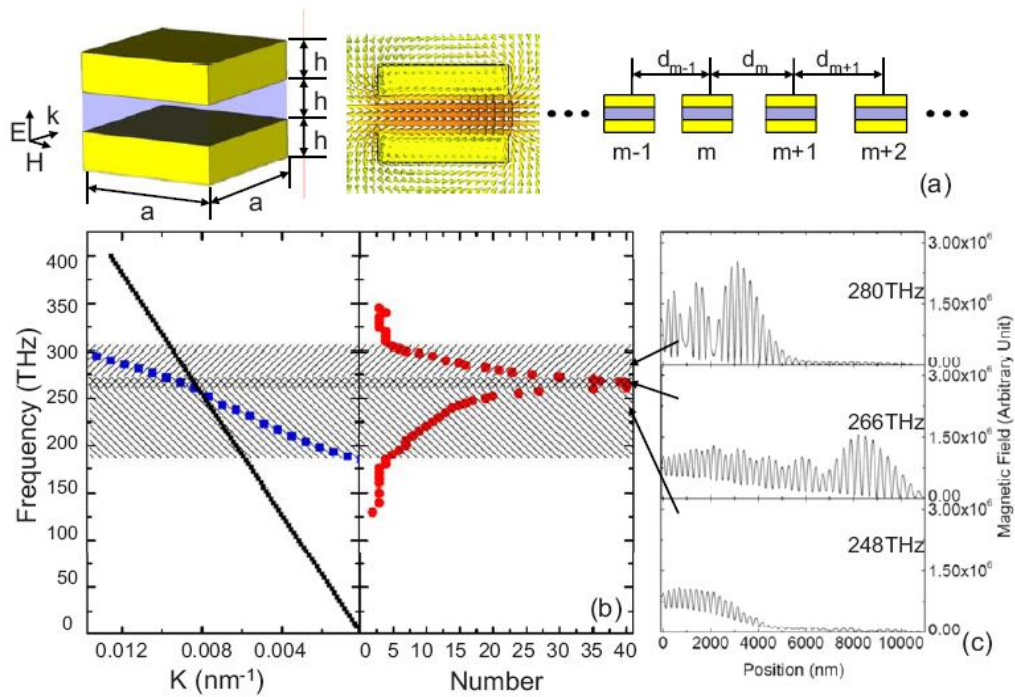

**Figure 11** (a) The geometry of a metallic trilayer structure, the magnetic field distribution at the magnetic resonant frequency, and the model of the graded nanosandwich chain; (b) the dispersion relation and the propagation length of the chain; (c) the magnetic field localizations at three different frequencies. From [53].

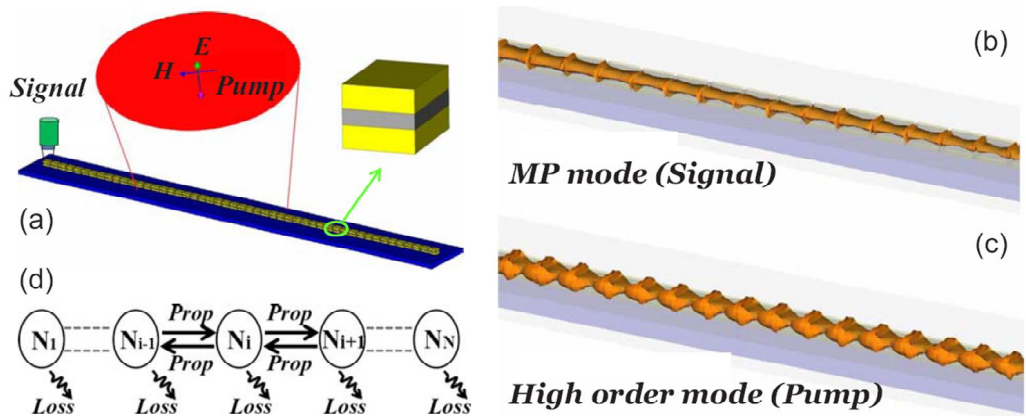

**Figure 12** (a) The geometry of the nanosandwich waveguide; (b) and (c) represent the MP mode and the high-order mode, respectively; (d) plot of the sketch of the coupled waveguide model. From [58].

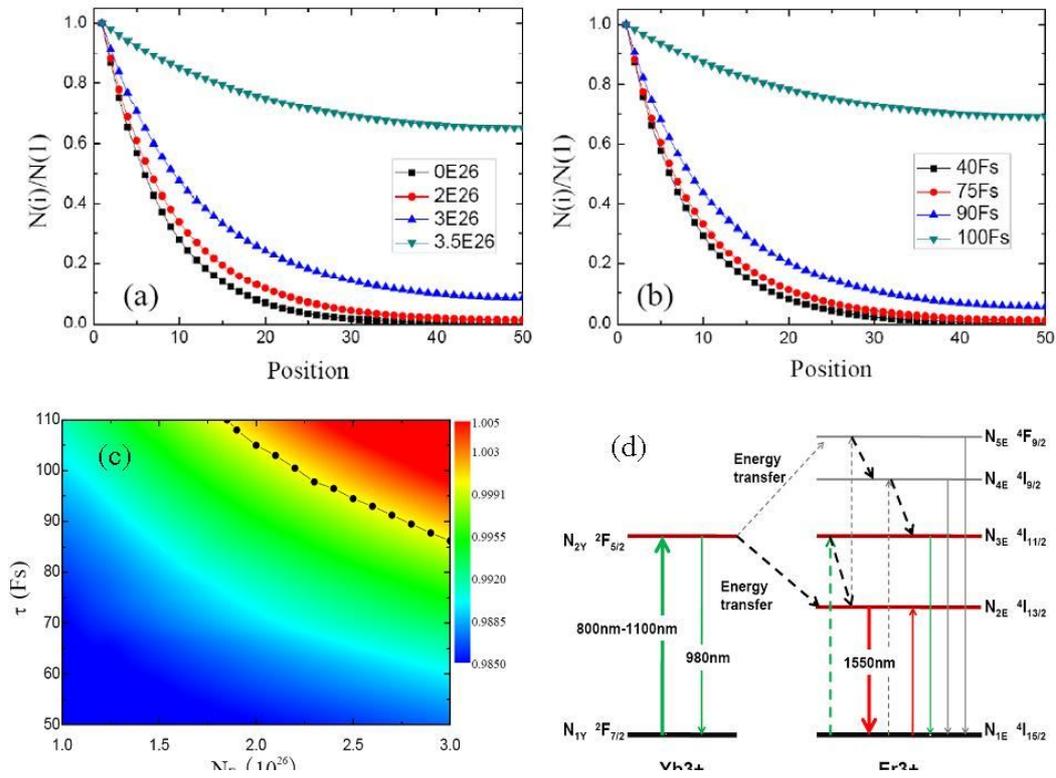

**Figure 13** The normalized numbers of photons in the nanosandwiches along the waveguide with different doping concentration $N_E$ and different decay time of signal $\tau$ are presented in (a) and (b), respectively. (c) The gain ability of single nanosandwich in the waveguide as the function of $Er^{3+}$ concentration $N_E$ and decay time $\tau$. The thresholds of amplification radiation are remarked by black dot-line curve. From [58].

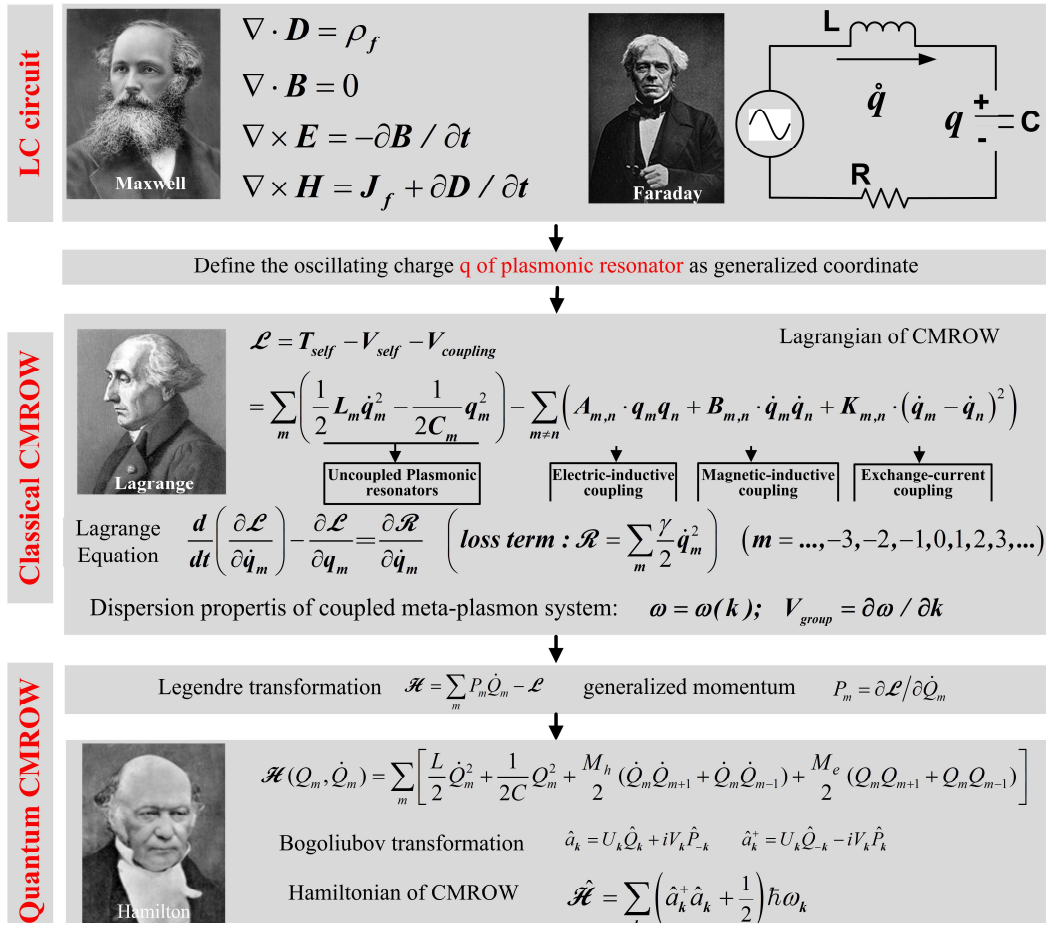

**Figure 14** From LC circuit and Lagrangian model to Hamiltonian model of CMROW.

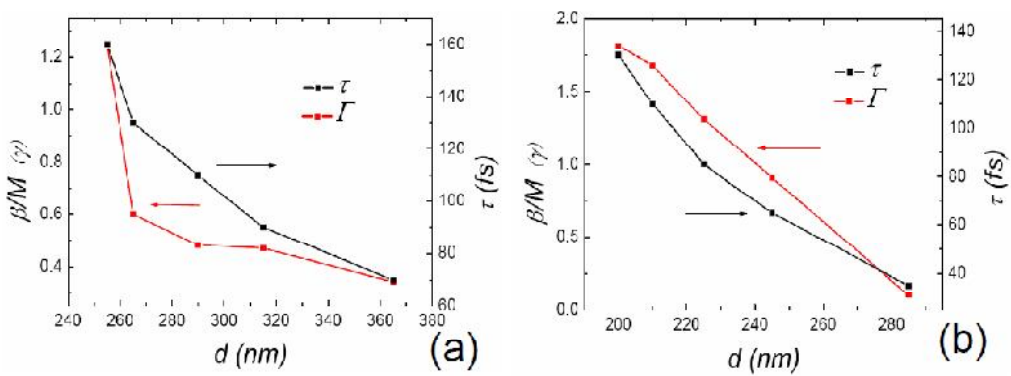

**Figure 15** (a) and (b) The stimulated emission coefficient and lifetime of the quasi-particle of CMROW in case I and case II with different spacings. From [84].